\documentclass[prd,
 onecolumn,
groupedaddress,
 amsmath,amssymb,
 aps,
 notitlepage,showpacs,showkeys,
]{revtex4-1}

\usepackage[utf8]{inputenc}
\usepackage[english]{babel}
\usepackage{graphicx,soul}
\soulregister\cite7
\soulregister\ref7
\usepackage{indentfirst}
\usepackage{amsmath,bm,dsfont,amssymb}
\usepackage{csquotes}
\usepackage{float,placeins}
\usepackage{wasysym}	
\usepackage{xcolor,transparent}
\usepackage[toc,page]{appendix}
\usepackage{import}

\usepackage[caption=false]{subfig}

\newcount\colveccount
\newcommand*\colvec[1]{
        \global\colveccount#1
        \begin{pmatrix}
        \colvecnext
}
\def\colvecnext#1{
        #1
        \global\advance\colveccount-1
        \ifnum\colveccount>0
                \\
                \expandafter\colvecnext
        \else
                \end{pmatrix}
        \fi
}

\newcommand{\vt}{\vartheta	}
\newcommand{\bs}{\boldsymbol}
\newcommand{\vp}{\varphi}
\newcommand{\de}{\partial}

\newcommand{\G}{\mathsf{G}}
\newcommand{\F}{\mathsf{F}}
\newcommand{\atan}{\mathrm{atan}}
\usepackage{hyperref}
\usepackage{url}

\begin{document}

\title{Mapping weak lensing distortions in the Kerr metric}

\author{Arianna I. Renzini}
 \email{arianna.renzini15@imperial.ac.uk}
\author{Carlo~R. Contaldi}
 \email{c.contaldi@imperial.ac.uk}
\author{Alan Heavens}
 \email{a.heavens@imperial.ac.uk}
\affiliation{
 Blackett Laboratory, Imperial College London, South Kensington Campus, London SW7 2AZ, United Kingdom
}

\date{\today}

\begin{abstract}
Einstein's theory of General Relativity implies that energy, i.e. matter, curves space-time and thus deforms lightlike geodesics, giving rise to gravitational lensing. This phenomenon is well understood in the case of the Schwarzschild metric, and has been accurately described in the past; however, lensing in the Kerr space-time has received less attention in the literature despite potential practical observational applications. In particular, lensing in such space is not expressible as the gradient of a scalar potential and as such is a source of curl-like signatures and an asymmetric shear pattern. In this paper, we develop a differentiable lensing map in the Kerr metric, reworking and extending previous approaches. By using standard tools of weak gravitational lensing, we isolate and quantify the distortion that is uniquely induced by the presence of angular momentum in the metric. We apply this framework to the distortion induced by a Kerr-like foreground object on a distribution of background of sources. We verify that the new unique lensing signature is orders of magnitude below current observational bounds for a range of lens configurations.
\end{abstract}


\maketitle
\section{Introduction}\label{sec:Intro}
It is well known that light traveling through space is bent according to the mass distribution it encounters and is thus gravitationally lensed. The study of lensing is key to mapping out the true positions of sources and infers the distribution nature of matter in our Universe; it has been a thriving field of research in the past decades, under the purely mathematical \cite{Iyer2006}, astrophysical \cite{Stebbins1996,Bartelmann2001}, and cosmological \cite{Munshi2008,Heavens2011,Kilbinger2015} points of view.

Gravitational lensing is typically studied in two limits, depending on whether the effect is \textit{weak}, where it is often analyzed as a statistical effect, or \textit{strong}. Strong lensing is immediately recognizable as background objects appear heavily warped, up to the occurrence of an \textit{Einstein ring} around the lens given by a single light source \cite{Bartelmann2010}. Weak lensing effects, on the other hand, may be revealed only via statistical data analysis, which investigates the nature of the mass distribution between the observer and lens. A milestone of weak lensing theory is the understanding of cosmic shear, which has proven to be a precious tool for the reconstruction of the dark matter distribution \cite{Kaiser1993} and which may be effectively modeled on the whole sky via tensor spherical harmonics \cite{Stebbins1996}. Weak lensing surveys in the last 20 years have proven more and more fruitful, starting from the 2000 Canada France Hawaii Telescope (CFHT) survey, which first detected a weak gravitational lensing signal by large-scale structure \cite{VanWaerbeke2003}, and the Red-Sequence Cluster Survey in 2001, which made use of both the CFHT and the Cerro Tololo Inter-American Observatory 4 m telescope \cite{Hoekstra2001}. Future data are equally promising, as shown by the first results of the Kilo-Degree Survey \cite{KiDS2015}, which started in 2011 at the European Southern Observatory's Very Large Telescope in the Atacama Desert, Chile, and those from the Dark Energy Survey \cite{Becker2015,DarkEnergySurveyCollaboration2016}, which began in 2013, and by  the Euclid assessment study report \cite{Laureijs2009}. As the vast majority of matter in the Universe is dark, the study of weak lensing is a necessary reality check when probing new matter and/or gravitational theories; for example, the analysis of galaxy-galaxy lensing around individual galaxies and clusters and bullet cluster reconstructions of the dark matter distribution in individual lenses have had fundamental implications on the nature of dark matter and models of modified gravity \cite{Clowe2006}.

The statistical treatment of survey data involves estimating the correlation in the ellipticity of galaxy images \cite{Kaiser1993}, that are distorted by the spin-2 shear field induced by lensing. These may be sourced by both scalar and nonscalar features of the matter distribution, where the latter is induced by dynamical properties. The consequence of any nonscalar property of the matter distribution is the appearance of a curl-like mode in the resulting lensing. In cosmic shear surveys gradient and curl-like components of the distortion are separated into $E$ and $B$ modes, where $E$ modes are parity conserving and $B$modes are parity violating \cite{Munshi2008}. The presence of $B$ modes is used as a standard test of the validity of the data and/or may be viewed as the signal of ``beyond-linear'' aspects of the lens distribution. Indeed, $B$ modes are expected to be induced by non linear structure formation resulting in intrinsic alignment of spatially correlated lenses. This signature is related to the vorticity of structure formation sourced by nonlinear evolution \cite{Hu1997}. At the level of individual lenses, we expect any angular momentum of the lens to be a source of parity violation, as it defines a preferred direction and breaks spherical symmetry; the presence of a rotation axis introduces handedness and thus prohibits left-right symmetry with respect to the axis.

The case of the Kerr metric is in this sense particularly interesting, as it is the simplest example of space-time which generates a nontrivial antisymmetric signature on a lensed background. This may be interpreted as an overall parity-violating lensing effect. Lensing in the Kerr metric has been explored considerably in the past, primarily as a mathematical challenge \cite{Boyer1967,Iyer2009,Aazami2011,Aazami2011a}, but also as an exercise in numerical relativity \cite{James2015}. That being said, our principal motivation for this work is to successfully express and quantify the antisymmetric component of the linear distortion in the weak lensing regime in the Kerr space-time, which is sourced uniquely by the angular momentum of a rotating lens. We do not expect this signal to be sufficiently large to be discernible from other lensing effects plus noise by present or near-future experiments; however, it is worth estimating how small the effect is likely to be for realistic astrophysical sources, as advanced surveys may be able to detect it. Weak lensing approximations used throughout are such that our analysis cannot be extended to the strong lensing case.

This paper is organized as follows. In Sec.~\ref{ReLeSchwaMe}, we briefly review lensing in the Schwarzschild metric case and derive well-known expressions to be used as a baseline to build and validate our calculation in the Kerr case. In Sec.~\ref{sec:kerr}, we analyze light bending in the Kerr metric. We first study geodesics in the Kerr space-time and develop the geometric quantities required to describe the lensing. The aim is to obtain an expression for the  lensing as a function of observable angles. We then interpret our results in terms of a linear distortion operator relevant for describing the effects in the weak lensing regime.  We conclude with a summary and discussion of our results in Sec.~\ref{discussion}, including an estimation of the magnitude of the new effects in an astrophysical setting. 

\section{Review: Lensing in the Schwarzschild Metric}\label{ReLeSchwaMe}
The geometry of gravitational lensing in the Schwarzschild metric is well understood, and has been used throughout this work as a basis on which to build more general lensing corrections. We review briefly review the nature and interpretation of the deflection angle induced by a spherically symmetric, static metric. We will refer back to the geometric setup described here when we develop the calculation for the Kerr metric case in Sec.~\ref{sec:kerr}. The 
Schwarzschild metric describes the metric of the space-time surrounding a body of mass $M$ with the line element given by
\begin{equation}
ds^2 = -\left(1-\frac{2GM}{r c^2}\right)\,c^2\,dt^2 +\left(1-\frac{2GM}{r c^2}\right)^{-1}\,dr^2+r^2\left(d\theta^2+\sin^2\theta\,d\phi^2\right)\,,
\end{equation}
where $c$ is the speed of light in the vacuum; $G$ is Newton's gravitational constant; and we have time coordinate $t$, radial coordinate $r$, and spherical angular coordinates $\theta$ (colatitude) and $\phi$ (longitude). The Schwarzschild radius $r_s = \frac{2GM}{c^2}$ can be introduced as a useful scale, and the mass of the source can be rescaled into units of length by introducing $m=\frac{G\,M}{c^2} = \frac{r_s}{2}$ such that either $m$ or $r_s$ can be used as the single scale factor parametrizing the metric. As is conventional in this type of analysis we assume a flat geometry in the limit where the impact parameter is much larger than the Schwarzschild radius, i.e. in the asymptotic region around the source. 

The total deflection angle $\alpha$ induced on a particle trajectory as it travels through the space-time, as measured at infinity in the Schwarzschild metric, can be written \cite{Wald1984} as an integral over the radial coordinate $r$ as,
\begin{equation}
\alpha=2\int_{\infty}^{r_0}\left|\frac{\dot{\phi}}{\dot{r}}\right|dr-\pi\,,
\label{alphaint}
\end{equation}
where $r_0$ is the distance of closest approach and overdots denote derivatives with respect to the affine parameter $\lambda$ along the particle trajectory. The integrand can be evaluated using the equations of motion derived in the Schwarzschild space-time,
\begin{equation}
\left(\frac{\dot{\phi}}{\dot{r}}\right)^{-1}=r^2 \sqrt{\frac{1}{b^2}-\left(1-\frac{2m}{r}\right)\,\frac{1}{r^2}}\,,
\end{equation}
where $b$ is the impact parameter. The distance $r_0$ can be derived as a series expansion in $b$ by setting $\dot{r}=0$; one obtains
\begin{equation}
r_0=b\,\left[1-\frac{m}{b}-\frac{3}{2}\left(\frac{m}{b}\right)^2-4\left(\frac{m}{b}\right)^3-\frac{105}{8}\left(\frac{m}{b}\right)^4-48\left(\frac{m}{b}\right)^5+\mathcal{O}\left(\frac{m}{b}\right)^6\right]\,.
\label{rzero2}
\end{equation}
The choice of $\frac{m}{b}$ as an expansion parameter is quite natural: the impact parameter is assumed to be much larger than the Schwarzschild radius $2m$, i.e., the coordinate singularity of the metric. The integration in Eq. (\ref{alphaint}) is carried out by implementing the method outlined in Ref.\cite{Keeton2005}: the integrand is rewritten as a function of $x=\frac{r_0}{r}$ and the parameter $h=\frac{m}{r_0}$, integrated in $x$ and expanded in $h$. Rewriting the result as an expansion in $\frac{m}{b}$ one obtains 
\begin{equation}
\alpha=4 \left(\frac{m}{b}\right)+\frac{15 \pi}{4} \left(\frac{m}{b}\right)^2+\frac{128}{3} \left(\frac{m}{b}\right)^3+\frac{3465 \pi}{64}  \left(\frac{m}{b}\right)^4+\frac{3584}{5} \left(\frac{m}{b}\right)^5+\mathcal{O}\left(\frac{m}{b}\right)^6\,,
\label{alphaexp2}
\end{equation}
which is commonly known as the Schwarzschild series \cite{Iyer2009}.

\begin{figure}[ht]
\centering
\includegraphics[width=0.85\textwidth]{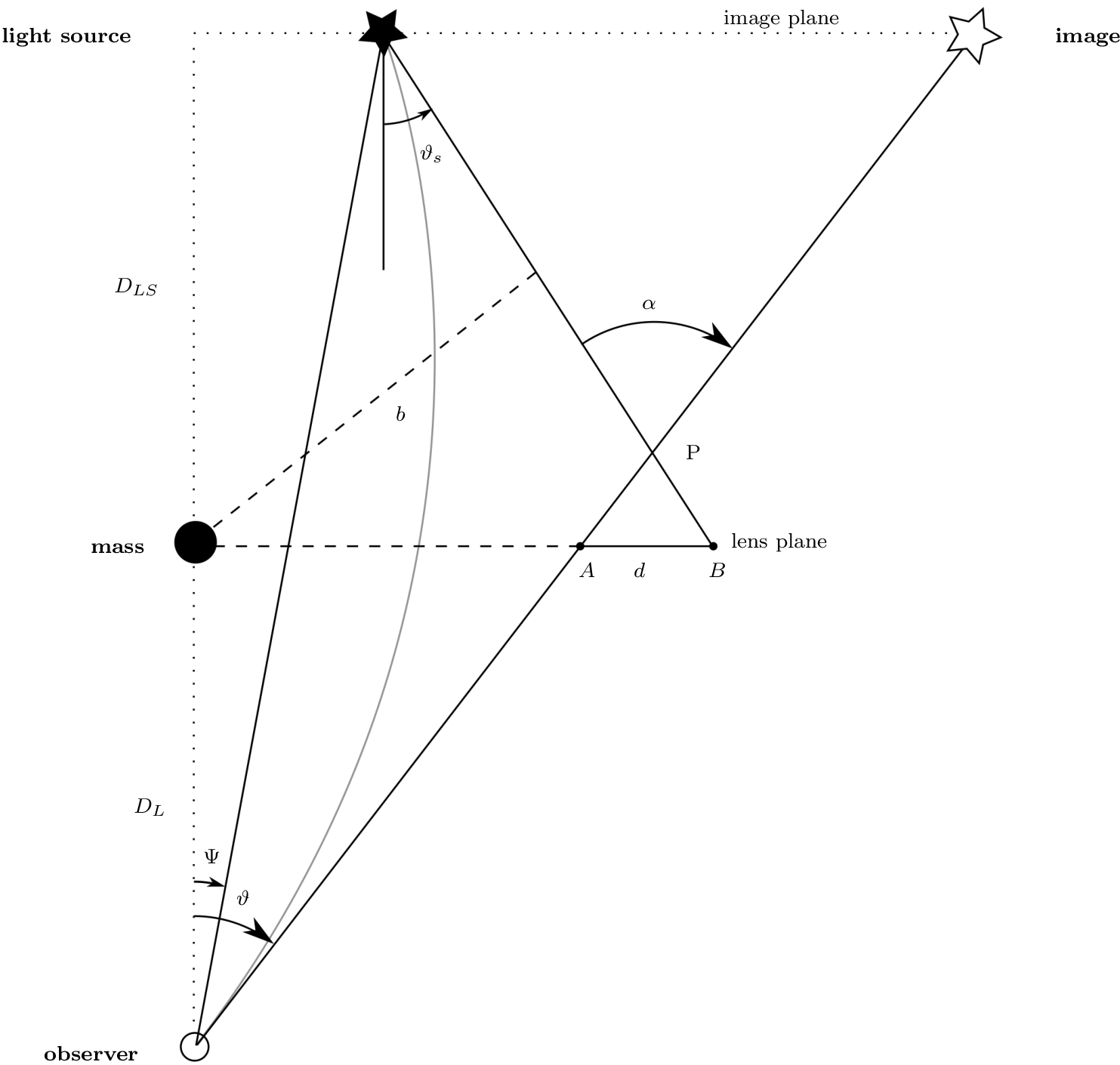}
\caption{View of the motion plane, i. e., the plane containing the light path, of lensing in the Schwarzschild space-time. The tangents to the motion intersect the lens plane in points $A$ and $B$, and $d$ is the distance between them. The observer receives a photon arriving along a distorted trajectory which gives the source an apparent angular position that is displaced from the true source position.}
\label{schwasym}
\end{figure}

In general, the source, lens, and observer are not aligned, and there will be a displacement $d$ between the intersections of the tangents to the ray at the observer and at the light source and the lens plane that has to be taken into account when carrying out beyond first-order calculations. The equation for $d$ may be read off from the plane containing the light path, which will be referred to as the motion plane,
\begin{equation}
d=q\left[(D_L+D_{LS}) \tan\Psi +D_{LS}\tan\vt_s-D_L\tan\vt \right]\,,
\label{d1}
\end{equation}
where $\Psi$ is the unknown angle spanned by the light source on the observer's sky, $D_L$ and $D_{LS}$ are the distances between the observer and lens and the lens and light source planes, respectively. In a cosmological context, these distances are angular diameter distances. $\vt$ and $\vt_s$ are the angles formed by the optical axis and the tangents to the ray at the observer and at the light source, respectively (see Fig.~\ref{schwasym}). The term $q$ is the overall sign which depends on what side of the lens plane the two tangents meet,
\begin{equation}
q=\begin{cases} +1,\,\,\,\vt>\vt_s\\ -1,\,\,\,\vt<\vt_s \end{cases}\,.
\end{equation}
For use below, we introduce the sum of angles as $\alpha=\vt+\vt_s$.

A lens equation will give $\Psi$ as a function of observable angles on the motion plane. Lens equations have been discussed much in the past, as it is not completely straightforward to handle angles and geometric objects in a curved space-time; an extensive review of the subject may be found in Ref.~\cite{Bozza2008}. We choose  lens equation for $\Psi$ that is obtained by simply applying Pythagoras' theorem, along with a number of considerations well motivated in Ref.~\cite{Bozza2008}:
\begin{equation}
D_S\,\tan\Psi=D_L\,\frac{\sin\vt}{\cos\vt_s}-D_{LS}\,\tan\vt_s\,.
\label{lensequ}
\end{equation}
Substituting Eq.~(\ref{lensequ}) in Eq.~(\ref{d1}), the displacement is given by
\begin{equation}
d=D_L\,\sin\vt\left[ \frac{1}{\cos\vt_s}-\frac{1}{\cos\vt} \right]\,,
\label{d2}
\end{equation}
as in Ref.~\cite{Aazami2011}. Note that, expanding to first order in $\vt$ and $\vt_s$, the displacement $d$ simply vanishes, and one recovers the typical weak lensing equation~\cite{Heavens2011}.\\
Having introduced most of the terminology and concepts with the Schwarzschild case, we now turn to the more involved scenario of the Kerr metric.

\section{Lensing in the Kerr Metric}\label{sec:kerr}
The Kerr metric is given by the line element
\begin{equation}
ds^2=-\frac{\Delta-a^2 \sin^2\theta}{\Sigma}\,dt^2+\frac{\Sigma}{\Delta}\,dr^2+\Sigma\,d\theta^2+\frac{(r^2+a^2)^2-\Delta\, a^2\sin^2\theta}{\Sigma}\,\sin^2\theta\,d\phi^2\\
+\frac{4\,m\,a\,r\,\sin^2\theta}{\Sigma}\,dt\,d\phi\,,
\label{Kerr}
\end{equation}
with 
\begin{equation}
\Delta(r)\equiv r^2-2\,m\,r+a^2\,\,\,,\quad \Sigma(r,\theta)\equiv r^2+a^2\,\cos^2\theta\,.
\end{equation}
It belongs to the two-parameter family of stationary axisymmetric metrics \cite{Visser2007}. In the expression above, the two chosen parameters are the mass parameter $m$ as given in Sec.~\ref{ReLeSchwaMe} and the mass-scaled angular momentum $a=\frac{J}{c\,M}$, measured at infinity. Note that both of these parameters have dimensions of length. The coordinates $(t,\,r,\,\theta,\,\phi)$ are known as \textit{Boyer-Lindquist} coordinates \cite{Boyer1967}, where $\theta$ and $\phi $ are the polar and azimuthal angles respectively, defined with respect to the rotation axis of the mass. These coordinates differ from those introduced in the Schwarzschild case in their relation to Cartesian coordinates,
\begin{eqnarray}
x&=&\sqrt{r^2+a^2}\,\sin\theta\,\cos\phi\,,\nonumber\\
y&=&\sqrt{r^2+a^2}\,\sin\theta\,\sin\phi\,,\\
z&=&r\,\cos\theta\,.\nonumber
\end{eqnarray}

It is convenient to transform $\theta$ to $\zeta=\frac{\pi}{2}-\theta$ and rewrite the metric in the modified coordinates $(t,\,r,\,\zeta,\,\phi)$, where $\zeta\in[-\frac{\pi}{2},\frac{\pi}{2}]$, and work in the quasiequatorial regime. This corresponds to the case where the observer lies on the equatorial plane (which is uniquely defined in the Kerr space-time) and the light source is slightly off it.
The metric then takes the form
\begin{equation}
ds^2=-\frac{\Delta-a^2 \cos^2\zeta}{\Sigma}\,dt^2+\frac{\Sigma}{\Delta}\,dr^2+\Sigma\,d\zeta^2+\frac{(r^2+a^2)^2-\Delta\, a^2\cos^2\zeta}{\Sigma}\,\cos^2\zeta\,d\phi^2+\frac{4\,m\,a\,r\,\cos^2\zeta}{\Sigma}\,dt\,d\phi\,,
\label{newKerr}
\end{equation}
with
\begin{equation}
\Sigma(r,\zeta)\equiv r^2+a^2\,\sin^2\zeta\,.
\end{equation}

Geodesic equations for massless particles in this space-time are obtained,
\begin{eqnarray}
\hat{\dot{t}}=&1+\frac{2m\,r\,(a^2-a\,\hat{L}+r^2)}{\Delta(r)\Sigma(r,\zeta)}\,,\\
\hat{\dot{r}}=&\pm\frac{\sqrt{r^4-(\hat{Q}+\hat{L}^2-a^2)r^2+2m\,[(\hat{L}-a)^2+\hat{Q}]r-a^2\,\hat{Q}}}{\Sigma(r,\zeta)}\,,\label{kerreoms}\\
\hat{\dot{\zeta}}=&\pm\frac{\sqrt{\hat{Q}+a^2\sin^2\zeta-\hat{L}^2\,\tan^2\zeta}}{\Sigma(r,\zeta)}\,,\\
\hat{\dot{\phi}}=&\frac{2\,a\,m\,r+\hat{L}(r^2-2m\,r)\,\sec^2\zeta+a^2\hat{L}\tan^2\zeta}{\Delta(r)\,\Sigma(r,\zeta)}\,,
\end{eqnarray}
where $\hat{L}$ and $\hat{Q}$ are two constants of the motion, rescaled with respect to the third constant of motion, the energy $E$, as
\begin{equation}\hat{L}=\frac{L}{E}\,\,\,,\qquad \hat{Q}=\frac{Q}{E^2}\,\,\,.\end{equation}
$L$ is the angular momentum, and $Q$ is the Carter constant \cite{Carter1968}. In the above, the hatted-dotting notation indicates derivatives with respect to the rescaled affine parameter $\hat\lambda=E\,\lambda$, such that, for example,  $\hat{\dot{r}}=\frac{d\,r}{d\,\hat\lambda}$.

From the equations of motion, bend angles may be calculated and written as functions of observable angles. It is important to note that the bend angle does not necessarily constitute a geometric angle in this setup, as in general the tangents to the motion at the observer and at the light source do not intersect in space. Thus, the bend angle $\alpha$ must be examined through its components on chosen planes. The most natural choice is to project it onto the equatorial (horizontal) and vertical planes; we will refer to these projections as $\alpha_{\rm hor}$ and $\alpha_{\rm ver}$, respectively.

\subsection{Geometry}
\begin{figure}[th]
\centering
\includegraphics[width=\textwidth]{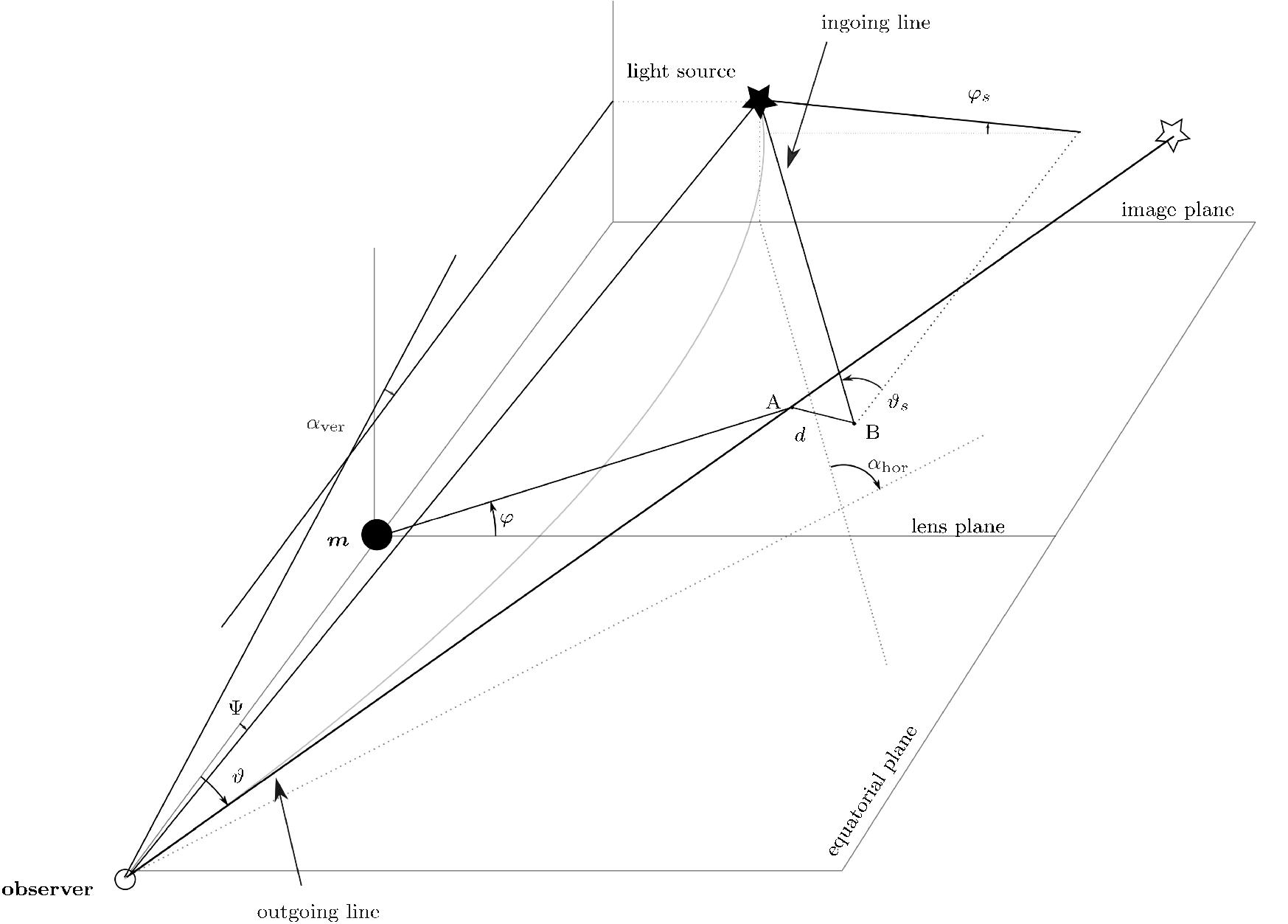}
\caption{$3D$ view of lensing in the Kerr metric. To draw this diagram, it is essential for the geometry to be asymptotically flat, and we may assume $b\gg m$ such that we are measuring angles in the flat region. The tangents to the motion intersect the lens plane in points $A$ and $B$.}
\label{kerr3d}
\end{figure}

\begin{figure}[th]
\centering
\includegraphics[width=0.7\textwidth]{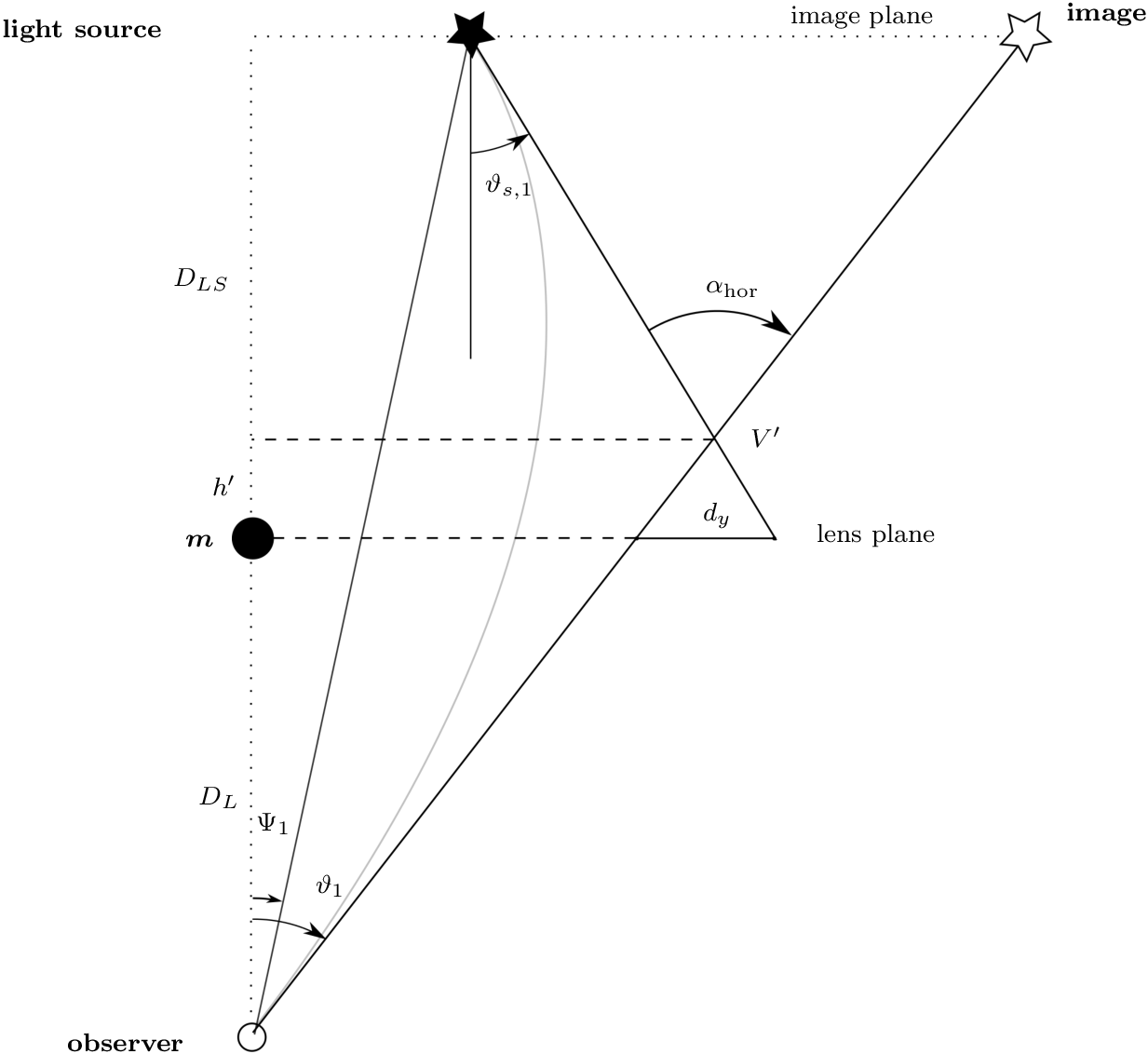}
\caption{View of the equatorial plane (i.e. the $(xy)$ plane) of our set up.  Note the \textit{displacement triangle} $(d_y,V')$ of base $d_y$ and vertex $V'$. $\vt_1$ and $\vt_{s,1}$ are the components on this plane of the angles $\bs\vt$, $\bs\vt_s$ respectively.}
\label{fig:kerrhor}
\end{figure}

\begin{figure}[t]
\centering
\includegraphics[width=0.67\textwidth]{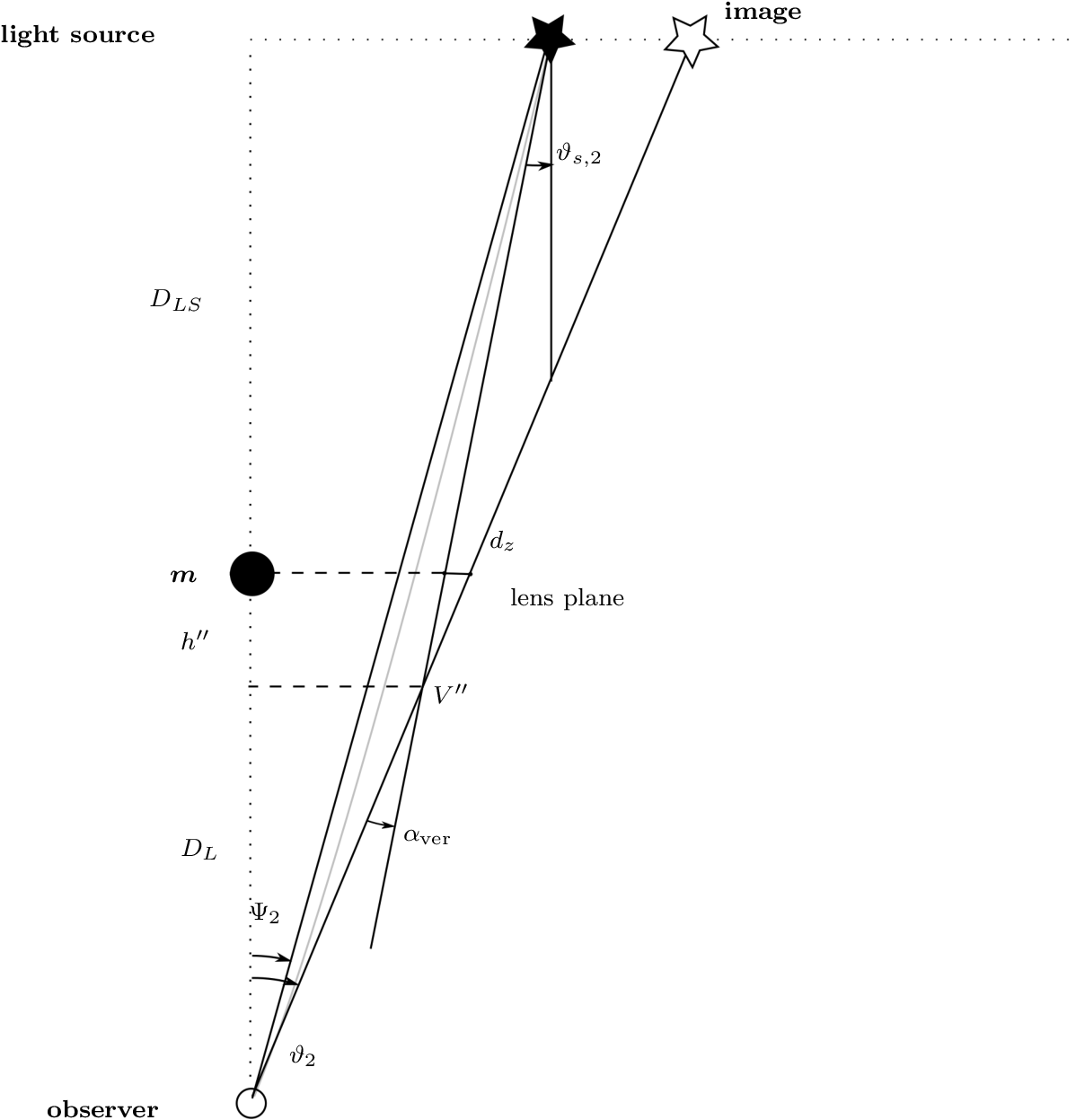}
\caption{ View of the vertical plane [i.e., the $(xz)$ plane] of our setup. Note the \textit{displacement triangle} $(d_z,V'')$ of base $d_z$ and vertex $V''$. $\vt_2$ and $\vt_{s,2}$ are the components on this plane of the angles $\bs\vt$ and $\bs\vt_s$, respectively.}
\label{fig:kerrver}
\end{figure}

The lensing geometry in the Kerr space-time may be derived in a way similar to the Schwarzschild case, with the complication that a photon's trajectory does not lie on a single plane. Thus, it is necessary to describe geodesics in full 3D space, as shown in Fig.~\ref{kerr3d}; it is, however, useful to project the light rays onto equatorial and vertical planes and derive expressions for the projected angles shown in Figs. \ref{fig:kerrhor} and \ref{fig:kerrver}. Expressions for the vertical and horizontal components of the total bend angle $\alpha$ are obtained by following closely the approach of Aazami \textit{et al.} \cite{Aazami2011, Aazami2011a}, with some meaningful differences which will be key to the development of a viable lensing map.

It is important to familiarize oneself with the observable angles $\vt$ and $\vp$ outlined in Figs. \ref{kerr3d}, \ref{fig:kerrhor}, and \ref{fig:kerrver}: $\vt$ is the angle between the tangent to the ray's trajectory at the observer and the optical axis, and $\vp$ is the rise of this line off the equatorial plane. The components of $\vt$ on the horizontal and vertical planes, $\vt_1$ and $\vt_2$, respectively, are the coordinates used on the image plane. The angle $\vp$ is then simply related to these via 
\begin{equation}
\tan\vp=\frac{\tan\vt_2}{\tan\vt_1}
\end{equation}
and will be used throughout as expansions around $\vp\sim0$ will be carried out to obtain the quasiequatorial regime. In the latter, the constants of motion which appear in (\ref{kerreoms}) may be written as \cite{Aazami2011} 
\begin{equation}
\hat{L}=s\,b\cos\vp\,\,,\qquad \hat{Q}=b^2\,\sin^2\vp\,,
\label{LQqe}
\end{equation} 
where $b\equiv D_L\,\sin\vt$ is the impact parameter just as in the Schwarzschild case. $\mathsf{s}$ is the sign of the angular momentum, and is either $+1$ for prograde motion, or $-1$ for retrograde motion. The tangent to the motion at the observer is referred to as the outgoing line and intersects the lens plane in point $A$; the tangent to the motion at the light source is referred to as the ingoing line and intersects the lens plane in point $B$.

The angles of interest are as follows:
\begin{itemize}
\item{\textbf{Image position:} The position of the image is fixed by the two angles $(\vt,\vp)$. $\bs\vt$ is the 2-dimensional angle subtended by the outgoing line and the optical axis $x$ and $\vt$ is its amplitude, $\vp$ is the angular position of $A$ on the lens plane. We choose their domains carefully: we set $0<\vt<\frac{\pi}{2}$, which spans only half of the image plane, but we can easily recover the other half by flipping the sign of the angular momentum $J$. The same goes for $0<\vp<\frac{\pi}{2}$. $\bs\vt$ has components on the horizontal and vertical planes $\vt_1$, $\vt_2$ respectively which are related to $(\vt,\vp)$ via
\begin{equation}
 \tan\vt_{1}=\tan\vt\cos\vp\,,\qquad  \tan\vt_2=\tan\vt\sin\vp\,.
\label{thetaones}
\end{equation}
}

\item{\textbf{Source position:} The position of the source is pinpointed by the angles $(\vt_s,\vp_s)$, which have the domains: $-\frac{\pi}{2}<\vt_s<\frac{\pi}{2}$; $0<\vp_s<2\pi$. $\vt_s$ quantifies the deviation of the incoming line with respect to the optical axis direction and is the amplitude of $\bs\vt_s$, $\vp_s$ quantifies the deviation of the incoming line with respect to the equatorial plane. Effectively, the quasi-equatorial regime implies that $\vp_s$ is small. $\bs\vt_s$ has components on the horizontal and vertical planes $\vt_{s,1}$, $\vt_{s,2}$ respectively which are related to $(\vt_s,\vp_s)$ via

\begin{equation}
\tan\vt_{s,1}=\tan\vt_s\cos\vp_s\,,\qquad \tan\vt_{s,2}=\tan\vt_{s}\sin\vp_s\,.
\end{equation}
The position of the source with respect to the observer may also be described by the 2-dimensional angle $\bm\Psi$ subtended by the straight line connecting source and observer and the optical axis, which we will make use of when developing the distortion map.}
\item{\textbf{Bend angle:} Since the motion does not occur on a plane, there is no physical angle $\alpha$ as in the Schwarzschild case. Thus, we must project the incoming and outgoing lines on the equatorial plane $(xy)$ and vertical plane $(xz)$, where they form the angles $\alpha_{\rm hor}$ and $\alpha_{\rm ver}$ respectively.}
\end{itemize}

\subsubsection{Horizontal bend angle}

Let us first consider the projection of the total bend angle onto the equatorial plane, $\alpha_{hor}$, the derivation of which is analogous to that of $\alpha$ for a Schwarzschild lens; in fact, Eq.~(\ref{alphaint}) does not depend on the metric and may be used in the Kerr case to quantify light bending on the equatorial plane. Using Eq.~(\ref{kerreoms}), one obtains
\begin{equation}
\frac{\dot{\phi}}{\dot{r}}=\pm\frac{2 a m r+\mathsf{s}\,b r\,(-2 m+r) \cos\vp \sec^2\zeta+\mathsf{s}\,a^2 b  \cos\vp\, \tan^2\zeta}{\Delta(r) \sqrt{r \left(b^2 (2 m-r)+r^3+a^2 (2 m+r)\right)-a b \left(a b \sin\vp^2+\mathsf{s}\,4 m r  \cos\vp\right)}}\,.
\end{equation}
This may be simplified in the quasiequatorial regime by expanding to first order in $\zeta\sim0$ and considering $\vp$ and $a$ to be small so that second order terms $a\vp^2$ and mixed terms $a\vp\zeta$ may be neglected. However, it is important to maintain leading second-order terms in $\vp$, as the map needs to be differentiable to obtain distortion terms.
Thus, one obtains
\begin{equation}
\frac{\dot{\phi}}{\dot{r}}=\frac{r^{1/2}\,(\mathsf{s}\,2\,a\,m-2\,b\,m\,\cos\vp+b\,r\,\cos\vp)}{\Delta\,\sqrt{r^3+b^2\,(2\,m\,\mathsf{F}^2-\mathsf{G}\,r)}}\,\,\,,
\label{phioverrdot}
\end{equation}
where
\begin{equation}
\mathsf{F}\equiv 1-\mathsf{s}\frac{a}{b}\,\,\,, \qquad \mathsf{G}\equiv 1-\frac{a^2}{b^2}\,\,\,.
\label{FG}
\end{equation}
Note how the terms involving $\cos\zeta$ and $\tan\zeta$ do not contribute at first order, so the deflection on the equatorial plane only depends on the position $r$, with a correction $\cos\vp$ given by the rise off the plane.\\
The relationship between $b$ and $r_0$ is obtained by solving the equation of motion for $r$ in Eq.~(\ref{kerreoms}), setting the velocity to $0$; it is a cubic equation in $r_0$ and has a single real solution,

\begin{equation}
r_0=\frac{2\,b}{\sqrt{3}}\,\sqrt{\G}\,\cos\left[\frac{1}{3}\,\cos^{-1}\left(-3^{3/2}\,\frac{\mathsf{F}^2}{\mathsf{G}^{3/2}}\frac{m}{b}\right)\right]\,\,\,,
\label{rzerob}
\end{equation}
which Taylor expanded in $\frac{m}{b}\ll1$ yields

\begin{equation}
r_0=b\left[ \sqrt{\G}-\frac{\F^2}{\G}\left(\frac{m}{b}\right)-\frac{3\,\F^4}{2\,\G^{5/2}}\left(\frac{m}{b}\right)^2-\frac{4\,\F^6}{\G^4}\left(\frac{m}{b}\right)^3-\frac{105\,\F^8}{8\,\G^{11/2}}\left(\frac{m}{b}\right)^4+\mathcal{O}\left(\frac{m}{b}\right)^5 \right]\,\,\,.
\label{rzerobTaylor}
\end{equation} 
The integration of Eq.~(\ref{phioverrdot}) is performed by changing the variable to $x=\frac{r_0}{r}$ and rewriting the expression as a function of $x$ and $h=\frac{m}{r_0}$, so

\begin{equation}
\alpha_{\rm hor}=2\,\int_0^1\frac{2\,a\,h\,x+2\,b\,h\,x\cos\vp-b\cos\vp}{\left(1-2\,h\,x+\left(\frac{a}{m}\right)^2\,h^2\,x^2\right)\,\sqrt{\mathsf{G}\,(1-x^2)-2\,\mathsf{F}^2\,h\,(1-x^3)}}\,dx-\pi\,.
\end{equation}
Keeping $\F$ and $\G$ as implicit functions of $b$ for the time being and expanding the integrand in small $h$ as in the Schwarzschild case, an expression for $\alpha_{\rm hor}$ as an expansion in $h$ is obtained
\begin{equation}
\alpha_{\rm hor}=\left[\mathsf{a_0}\,\pi+4\,\mathsf{a}_1\,h-\left(4\mathsf{a}_{2,1}+\pi\,\mathsf{a_{2,2}}\right)\,h^2+\mathcal{O}(h^3)\right]\,,
\label{alphah}
\end{equation}
where
\begin{eqnarray}
\mathsf{a_0}&=& \frac{\cos\vp}{\sqrt{\G}}-1\,,\\
\mathsf{a_1}&=& \frac{\F^2\cos\vp+\G-\F\G}{\G^{3/2}} \,,\\
\mathsf{a_{2,1}}&=& \frac{\F^2\,(\F^2\cos\vp+\G-\F\G)}{\G^{5/2}} \,,\\
\mathsf{a_{2,2}}&=&\frac{\G\,(3\,\F^2+2\,\G)(1-\F)+\cos\vp\,\left(\frac{15}{4}\,\F^4\,-\frac{1}{2}\left(\frac{a}{m}\right)^2\,\G^2\,\right)}{\G^{5/2}}\,.
\end{eqnarray}
Substituting now Eqs.~(\ref{rzerobTaylor}) and (\ref{FG}) into Eq.~(\ref{alphah}) and reexpanding in $\frac{m}{b}$, one finally gets
\begin{eqnarray}
\alpha_{\rm hor}&=& 4\,\cos\vp\, \left(\frac{m}{b}\right)+\left(\cos\vp\,\frac{15 \pi}{4}+4\,\mathsf{s}\,\frac{a}{m}\,(1-4\,\cos\vp) \right)\left(\frac{m}{b}\right)^2+\nonumber\\
&+&\left(\frac{128}{3}
+5\pi\,\mathsf{s}\,\frac{a}{m}\,(1-3\cos\vp)
+4\cos\vp\,\left(\frac{a}{m}\right)^2\right) \left(\frac{m}{b}\right)^3
+\mathcal{O}\left(\left(\frac{m}{b}\right)^4\right)\,,
\label{ahormb}
\end{eqnarray}
This result agrees with with the Schwarzschild case [see Eq.~(\ref{alphaexp2})], as the latter is recovered by simply setting $a=0$ and choosing the plane $\vp=0$ as the equatorial plane (a choice that one can and must make to recover the Schwarzschild series, as the angle $\alpha$ lies geometrically on such plane).

\subsubsection{Vertical bend angle}

\begin{figure}[t]
\centering
\includegraphics[width=0.55\textwidth]{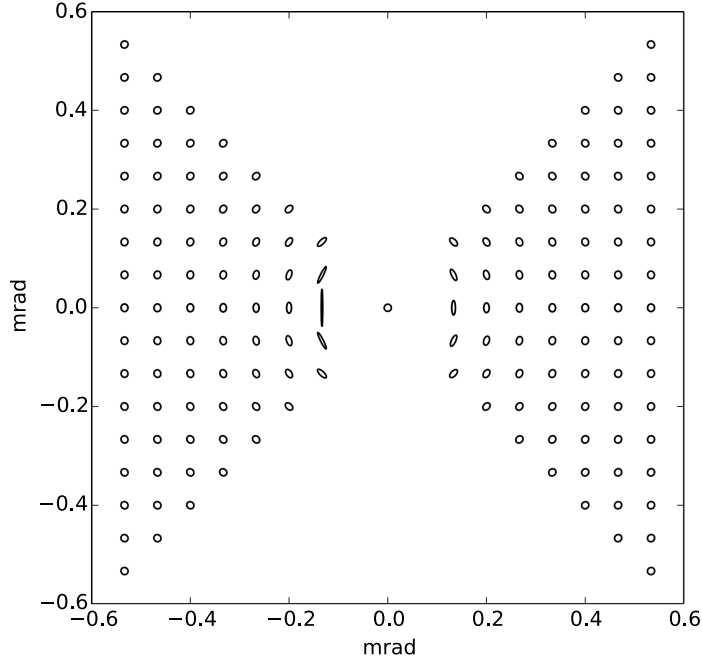}
\caption{We take a region of $1.2$ mrad$^2$ of the sky (which corresponds approximately to $5$ arc min$^2$) around a Kerr lens and plot the distortion brought upon a grid of circular sources. To produce a substantial effect, we input the values $M=5\cdot10^{12}\,M_{\astrosun}$, $\mathsf{a}=0.015$, where $\mathsf{a}$ is the rescaled rotation parameter: $\mathsf{a}=\frac{a}{R_g}$, with $R_g\sim100$ kpc. Motivation for this parameter choice may be found in Sec. \ref{discussion}, along with an explanation as to why this visualization is strictly unphysical. Note the asymmetry of the effect with respect to the $z$ axis. The visualization of the distortion closest to the point mass is unrealistic: for such values of $b$, one would expect also an overall arcing of the image, as in an Einstein ring; in fact, closer to the lens, the weak lensing approximation breaks down as $|\gamma|>1$. The values of the distances between observer and lens $D_L$ and lens and light source $D_{LS}$ are $ D_{L}=D_{LS}=100$ Mpc.}
\label{fig:kerrshear}
\end{figure}

Let us now consider the projection of the total bend angle onto the vertical plane, i. e., the plane perpendicular to the equatorial plane containing both the observer and the lens. Aazami \textit{et al.} \cite{Aazami2011a} obtained 
\begin{eqnarray}
\alpha_{\rm ver}&=&\vt_2-\atan\left(\tan\vp_s\,\tan(\alpha_{\rm hor}-\theta_1)\right)\nonumber\\
&&\approx\vt_2-\atan\left(\frac{\vp\,\sin\left[-2\,I(\infty) + \sin^{-1}\left(\sqrt\G\,\sin\vt\right)\right]}{\sqrt\G\,\cos\,(\alpha_{\rm hor}-\theta_1)}\right)\,,
\label{aver}
\end{eqnarray}
where $I(\infty)$ is the expansion in $\frac{m}{b}$,
\begin{equation}
I(\infty)=\frac{\pi}{2}+2\,\left(\frac{m}{b}\right)+\left(\frac{15\pi}{8}-4\,\mathsf{s}\,\frac{a}{m}\right)\,\left(\frac{m}{b}\right)^2+\left( \frac{64}{3}-\frac{15\pi\,\mathsf{s}\,a}{2\,m}+5\,\left(\frac{a}{m}\right)^2\right)\,\left(\frac{m}{b}\right)^3+\mathcal{O}\left(\frac{m}{b}\right)^4\,.
\end{equation}
This expression is first order in both $\vp$ and $\vp_s$, whereas the expression for $\alpha_{\rm hor}$ shown above is second order. However, this is not an issue as in the quasiequatorial regime $\alpha_{\rm ver}\ll\alpha_{\rm hor}$, and as such, the contribution of the former will be less appreciable.

\begin{figure}[t]
\centering
\includegraphics[width=0.52\textwidth]{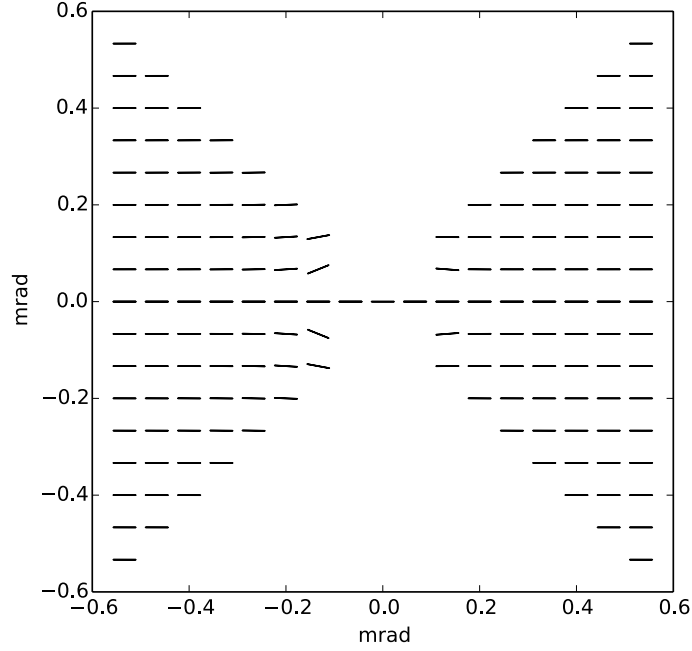}
\caption{We plot the effect of the $\rho$ field on a grid of vectors in a region of $1.2$ mrad$^2$ of the sky around a Kerr lens; there is no axial symmetry with respect to $z$, and as such, the behavior includes a $B$ mode. The values assumed for this plot are $M=5\cdot10^{12}\,M_{\astrosun}$, $\mathsf{a}=0.08$. The values of the distance between the observer and lens $D_L$ and the lens and light source $D_{LS}$ are $ D_{L}=D_{LS}=100$ Mpc.}
\label{fig:kerrvects}
\end{figure}

\subsubsection{Source position}
We now look at the true position of the lensed light source, $\Psi$, as a function of the observable angles introduced up to now. A careful geometrical analysis of the case yields the general lens equations with displacement 
\begin{eqnarray}
\tan\Psi_1&=&\tan\vt_1 - \left( \frac{D_{LS}}{
      h'} +\mathsf{q_y} \right)\,\frac{d_y}{D_S} \,,\\
\tan\Psi_2&=&\tan\vt_2 - \left( \frac{D_{LS}}{
      h''} + \mathsf{q_z} \right)\,\frac{d_z}{D_S} \,,
\label{lensquKerr}
\end{eqnarray}
where $d_y$, $d_z$ and $h'$, $h''$ are the bases and heights of the displacement triangles $(d_y,V')$ and $(d_z,V'')$,  respectively --- see Figs. \ref{fig:kerrhor} and \ref{fig:kerrver}. $d_y$, $d_z$ are derived in Ref.~\cite{Aazami2011}
\begin{equation}
d_y=D_L\,\sin\vt\,\cos\vp\,\left|\frac{1}{\cos\vt_s}-\frac{1}{\cos\vt}\right|\,,
\end{equation}
\begin{equation}
\begin{split}
d_z=&\biggl|-D_L\,\tan\vt\,\sin\vp+\frac{D_L\,\sin\vt}{1-\sin^2\vt_s\,\sin^2\vp_s}\times \\
&\left(\,\cos\vp\,\sin\vt_s\,\tan\vt_s\,\sin\vp_s\,\cos\vp_s+\sqrt{\sin^2\vp-\sin^2\vt_s\,\sin^2\vp_s} \,\right)\biggr|\,;
\end{split}
\label{dzkerr}
\end{equation}
$\mathsf{q_y}$ and $\mathsf{q_z}$ are the signs associated to $d_y$, $d_z$,
\begin{equation}
\mathsf{q_y}=\begin{cases} +1,\,\,\,\vt_1>\vt_{s,1}\\ -1,\,\,\,\vt_1<\vt_{s,1} \end{cases}\,\,\,, \qquad \mathsf{q_z}=\begin{cases} +1,\,\,\,\vt_2>\vt_{s,2}\\ -1,\,\,\,\vt_2<\vt_{s,2} \end{cases}\,.
\end{equation}
The heights $h'$ and $h''$ of the two displacement triangles are key in the lens equations (\ref{lensquKerr}). They are obtained by using simple Euclidean rules for scalene triangles, such as the following: given the angle $\beta$ and its opposite side $L$, the ratio $\frac{L}{\sin\beta}$ is constant throughout
the triangle, and $\sin\left(\frac{\pi}{2} - \beta \right) = \cos\beta$; one must also note that $\alpha_{\rm hor} = \vt_1+\vt_{s,1}$ and $\alpha_{\rm ver} = \vt_2+\vt_{s,2}$ just like in the Schwarzschild geometry. The heights are then
\begin{equation}
h'=d_y\,\frac{\cos\vt_1\, \cos\vt_{s,1}}{\sin\alpha_{\rm hor}}\,
\, ,\qquad
h''=d_z\,\frac{\cos\vt_2\, \cos\vt_{s,2}}{\sin\alpha_{\rm ver}}\,
\,.
\end{equation}

The source position $\bs\Psi=\colvec{2}{\Psi_1}{\Psi_2}$ may then be entirely written out as a function of the observable angle $\bs\vt=\colvec{2}{\vt_1}{\vt_2}$.

\begin{figure}[t]
\centering
\includegraphics[width=0.78\textwidth]{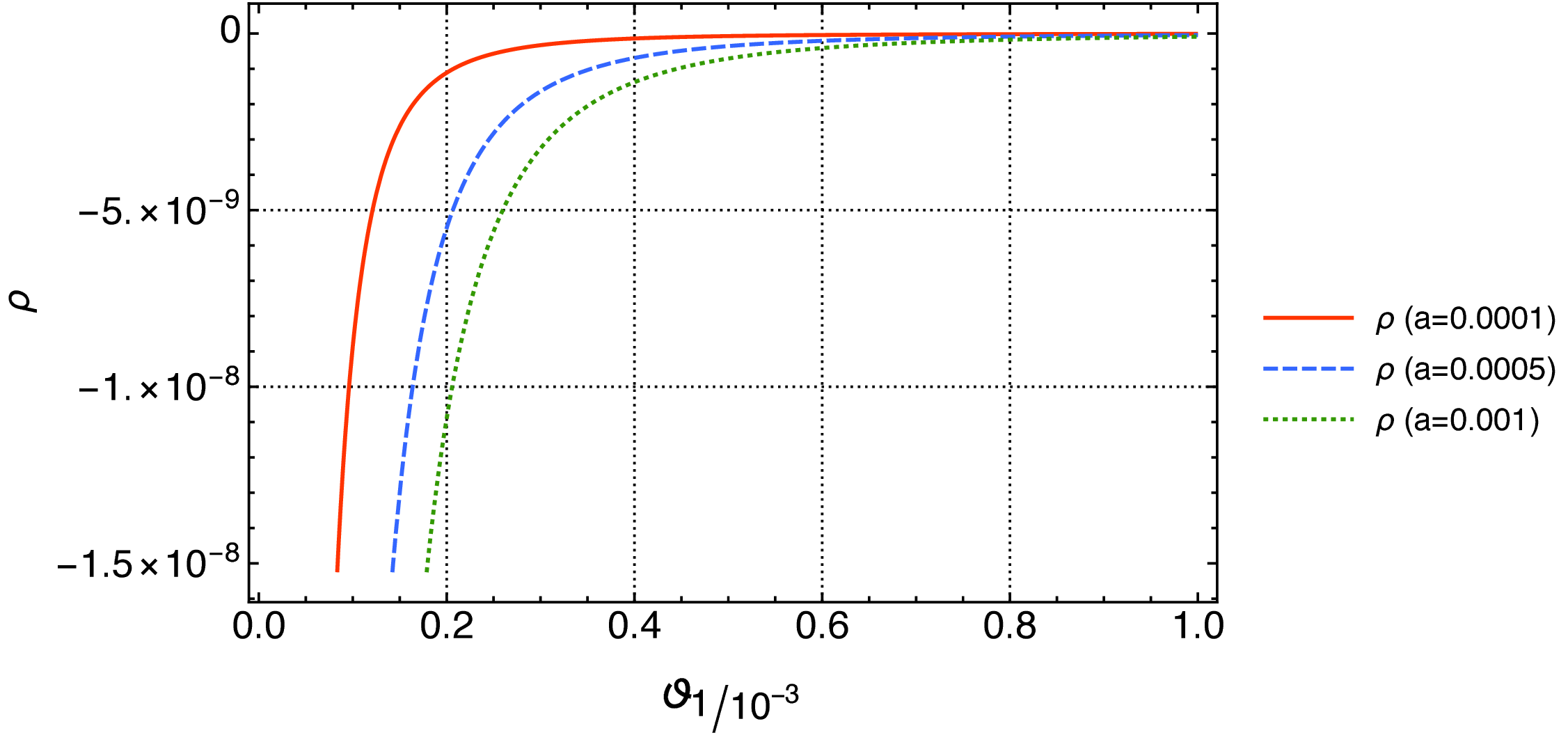}
  \caption{A plot of $\rho$ as a function of the horizontal component $\vt_1$ of the observed angle $\bs\vt$ at different values of the rescaled rotation parameter $\mathsf{a}$, keeping the impact and mass parameters fixed ($\vt=3\cdot10^{-4}$, $M=10^{12}\,M_{\astrosun}$). The chosen values of $\mathsf{a}$ may be considered possible physical values, as will emerge from the discussion in Sec. \ref{discussion}.}
\label{fig:rho_a}
\end{figure}

\begin{figure}[t]
\centering
\begin{minipage}{\textwidth}
\includegraphics[width=0.78\textwidth]{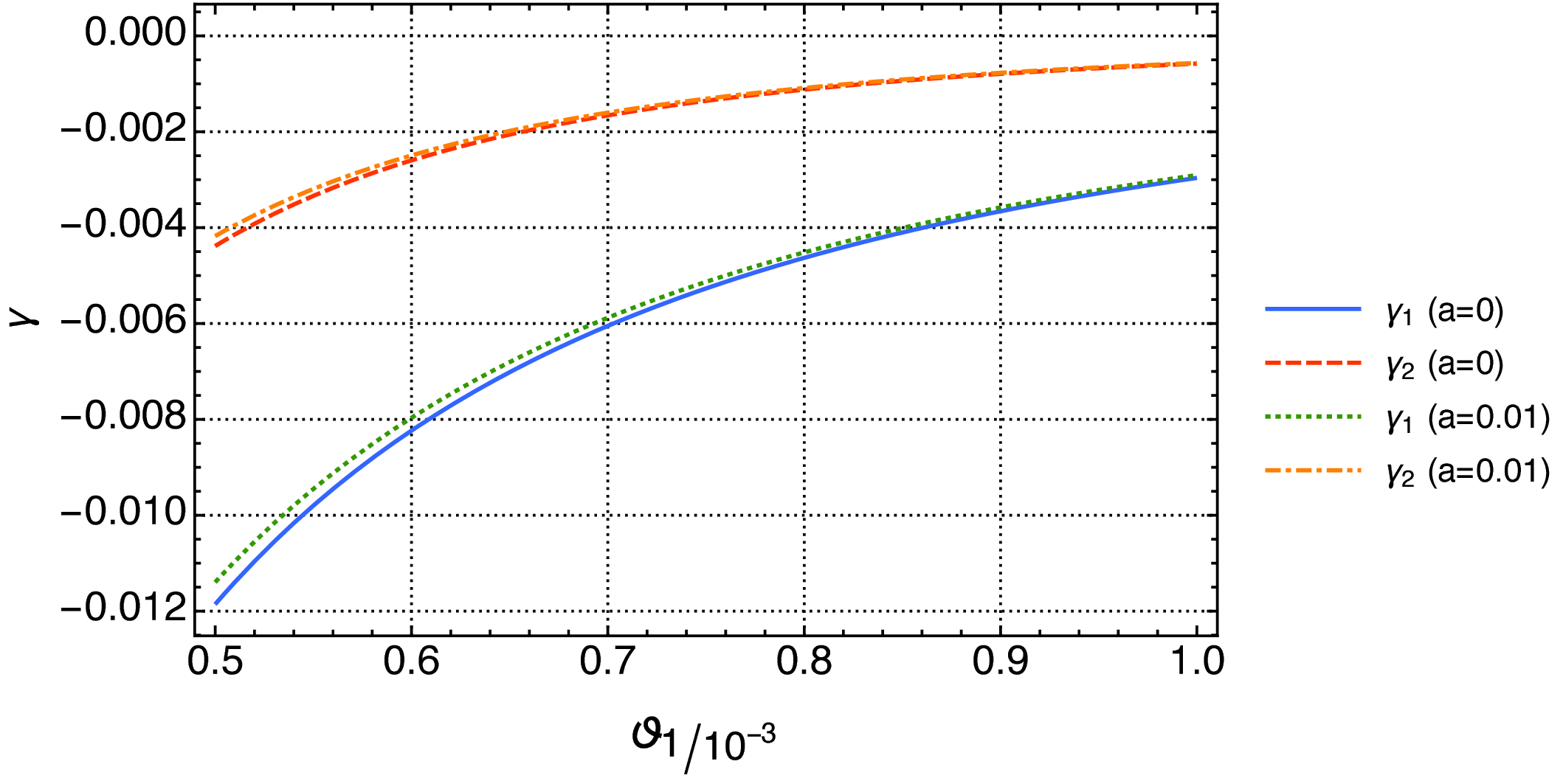}
\end{minipage}
\vskip 5mm
\begin{minipage}{\textwidth}
\centering
\includegraphics[width=0.78\textwidth]{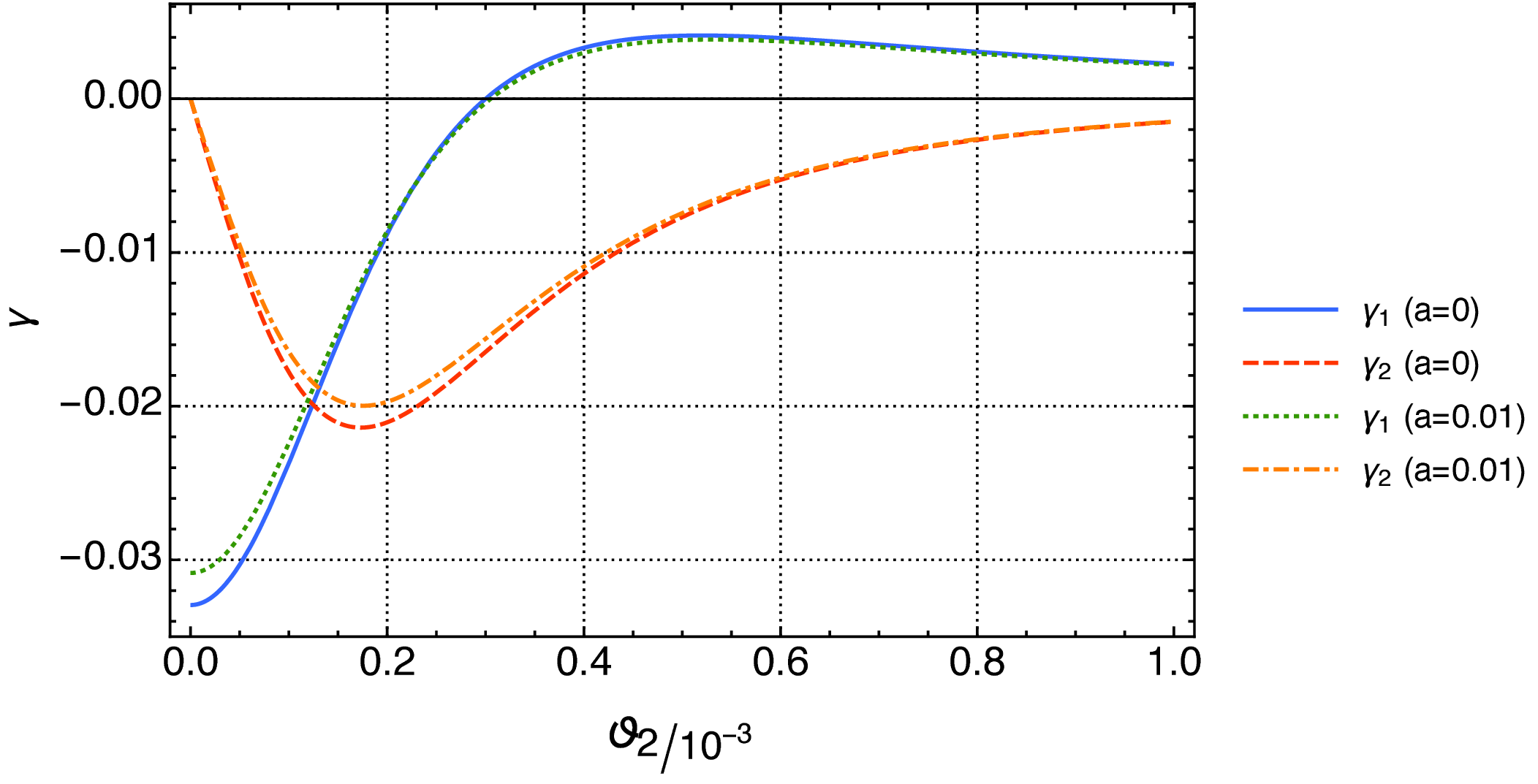}
\end{minipage}
\caption{Top: a comparative plot of the components of the shear field $\gamma_1$, $\gamma_2$ at fixed values of $\vt_2$, $m$ ($\vt_2=10^{-6}$ rad for $\gamma_1$, $\vt_2=10^{-3}$ rad for $\gamma_2$), varying $\vt_1$ respecting the region of validity, in the two cases $a=0$ (Schwarzschild) and $a\neq0$ (Kerr). Bottom: a plot of $\gamma_1$, $\gamma_2$ at fixed values of $\vt_1$, $m$ ($\vt_1=3\cdot10^{-4}$), varying $\vt_2$ respecting the region of validity, in the two cases $a=0$ and $a\neq0$. $M=10^{12}\,M_{\astrosun}$ for all lines. $\mathsf{a}$ is the rescaled rotation parameter for both plots. The choice of all parameter values was made bearing in mind the interpretive value of the plots.}
\label{fig:gammaonetwos}
\end{figure}

\subsection{Distortion map}
To study the distortion brought about by an object in the sky, we make use of a standard object used typically in weak lensing scenarios: the \textit{distortion matrix}, which is defined as \cite{Bartelmann2010, Munshi2008}
\begin{equation}
\mathcal{D}_{ij}=\frac{\de\Psi_i}{\de\vt_j}\,,
\label{Dij}
\end{equation}
where $\bm\Psi$ is the position of a light source and $\bm\vt$ is the position of its image with respect to the observer. $\mathcal{D}$ may be decomposed into trace and traceless parts; within this decomposition, three distinct fields may be identified, such that \cite{Thomas2009}
\begin{equation}
\mathcal{D}_{ij}=\begin{pmatrix}
    1-\kappa-\gamma_1       & -\gamma_2-\rho  \\
  -\gamma_2+\rho&   1-\kappa+\gamma_1      \\
\end{pmatrix}\,,
\label{Dwithrho}
\end{equation}
where $\kappa$, $\bs\gamma$ and $\rho$ are the convergence, shear, and rotation fields, respectively. The components of these fields may be extracted from the matrix
\begin{equation}
\gamma_1=\frac{1}{2}\,\left(\mathcal{D}_{22}-\mathcal{D}_{11}\right)\,, \qquad \gamma_2=-\frac{1}{2}\,\left(\mathcal{D}_{12}+\mathcal{D}_{21}\right)\,, \qquad \rho=\frac{1}{2}\,\left(\mathcal{D}_{21}-\mathcal{D}_{12}\right)\,.
\label{fields}
\end{equation}
These describe and quantify the distortion impressed on background sources by a foreground massive object. Extending this to the Kerr metric case, one may model the distorting effects brought about by a rotating object, with appropriate caveats \cite{Visser2007}.

We study the distortion in the simplest case: that of a field of circularly symmetric background sources. In the case of weak lensing, $\gamma\ll1$, a circular source with radius $R$ will be sheared into an ellipse with axes ($a$, $b$), given as 
\begin{equation}
a=R\,\sqrt{\frac{1+\gamma}{1-\gamma}}\,,\qquad b=R\,\sqrt{\frac{1-\gamma}{1+\gamma}}\,,
\label{ellipse}
\end{equation}
where 
$$\gamma=\sqrt{\gamma_1^2+\gamma_2^2}\,.$$
These expressions are easily derived assuming $\gamma\sim\epsilon$ ellipticity of the object \cite{Bartelmann2010} and considering the area of the image to be conserved. The resulting ellipse will also be rotated by a total angle $X\approx\omega+\rho/2$, where the angle $\omega$ is associated to the spin-2 shear field, 
\begin{equation}\omega=\frac{1}{2}\cos^{-1}\left(\frac{\gamma_1}{\gamma}\right)\,,\end{equation}
and $\rho$ is simply the magnitude of the rotation field \cite{Pen2006}.

   The elements of the distortion matrix are analytically derived by inserting the lens Eqs.~\ref{lensquKerr} into \ref{Dij}; note that the lens equations have been constructed so as to be differentiable in the angles of interest, and thus return nontrivial functions. By combining the elements of the distortion matrix as instructed in Eq.~\ref{fields}, one obtains analytic functions for the shear and rotation fields. Expressions for these functions are incredibly long and would give no insight at a glance, so we choose not to include them in the text. Carrying out this computation numerically in a manner as consistent and informative as the analytic one implemented here would have been very challenging, as the numerical precision needed is very high.\\ 
The overall distorting effects of the Kerr metric may be seen in Fig.~\ref{fig:kerrshear}. At a glance, the overall rotation effect appears symmetric around the source, yet this is not the case. The rotation field $\rho$ has, however, a much smaller impact than the shear field $\gamma$. The isolated effect of the $\rho$ field may be seen in Fig.~\ref{fig:kerrvects}, where we plot the rotation field on a grid of vectors around the lens, and in Fig.~\ref{fig:rho_a}, which presents the trend of $\rho$ at different values of the rotation parameter. Fig.~\ref{fig:gammaonetwos} shows comparative plots in the Schwarzschild and Kerr cases of the two components of the shear field, $\gamma_1$ and $\gamma_2$, so that the reader may appreciate the effect of rotation on the latter. For Figs.~\ref{fig:kerrshear}~--~\ref{fig:gammaonetwos} the values of the distances between the observer and lens $D_L$ and the lens and light source $D_{LS}$ are chosen to be $ D_{L}=D_{LS}=100$ Mpc. 


\begin{figure}[t]
\centering
\qquad\qquad
\includegraphics[width=0.8\textwidth]{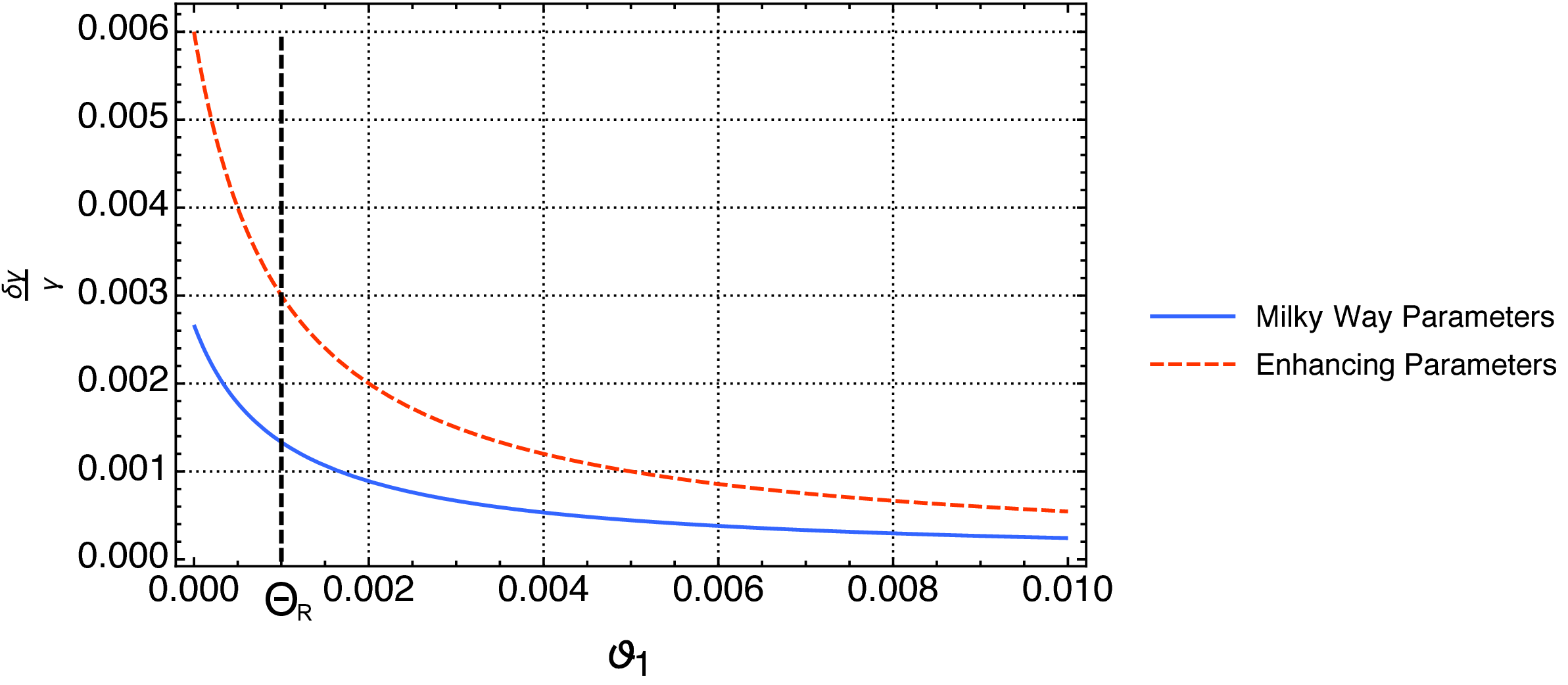}
\caption{A plot of the normalized shear difference $\frac{\delta\gamma}{\gamma}$ between the left and right shear [see Eq.~(\ref{deltagamma})] in two cases: for the Milky Way parameter set ($M_{\rm MW}=10^{12}\,M_{\astrosun}$, $R_{\rm MW}=100$ kpc and $\mathsf{a}_{\rm MW}=1.5\cdot10^{-3}$) and the set of parameters used to produced the enhanced effect depicted in Fig.~\ref{fig:kerrshear} ($M_{\rm Enh}=5\cdot10^{12}\,M_{\astrosun}$, $R_{\rm Enh}=100$ kpc and $\mathsf{a}_{\rm Enh}=1.5\cdot10^{-2}$), as a function of $\vt_1$. The plot is truncated at $\vt_1 = \Theta_R=0.1$ mrad $\approx10$ arc min for this parameter set. This truncation finds motivation in the discussion in Sec.~\ref{discussion}.}
\label{fig:deltagamma}
\end{figure}
\section{Astrophysical Implications}\label{discussion}

In the previous section, we developed a differentiable lensing map in the Kerr space-time, and obtained expressions for the shear and rotation fields.  A possible application of this map is extending the Kerr metric to represent rotating galaxies and estimate the distorting effects these may have on background radiation. One must be aware of the caveats of the case: as Birkhoff's theorem does not apply to the Kerr space-time, one may not assume that outside a rotating object the metric is exactly Kerr. However, we focus on weak lensing which occurs for rays traveling with an impact parameter $b$ much larger than the Schwarzschild radius $R_{Schw}$ associated to the object, and well outside the object itself. As the metric is asymptotically Kerr outside the rotating object \cite{Visser2007}, we conclude that the metric is viable if we restrict the analysis of the distortion to light rays with $b$ greater than the radius of the rotating galaxy $R_g$, as $R_g\gg R_{Schw}$. Also, one must consider that our Universe is best described by the Friedman-Robertson-Walker metric, whereas the Kerr metric is asymptotically flat (Minkowski). Nevertheless, we do not expect this to invalidate the model, as all distortion is imprinted to the light rays locally close to the lens and thus is insensitive to the expansion of the Universe.
    
A galaxy, unlike a black hole, is a large, spread-out object with $R_{\rm max}\gg R_{\rm Schw}$, where $R_{\rm Schw}=\frac{2\,M\,G}{c^2}\equiv2\,M_{\bullet}$; thus, its angular momentum may be much greater, at fixed mass. Considering the mass-scaled angular momentum $a$ as it is defined for the Kerr metric,
\[
a=\frac{J}{M_{\bullet}}=\frac{2\,J}{R_{\rm Schw}}\,,
\]
one finds that this is not a small parameter, and it makes no sense to look at first-order contributions to the bend angle in it. To use it properly and obtain an expansion parameter, one should scale it by the radius of the galaxy $R_g$; in fact, expanding $g_{tt}$ of the Kerr metric in $\frac{1}{R}$ and restricting it to the equatorial plane ($\theta=\frac{\pi}{2}$) for simplicity,
\begin{equation}
g_{tt}=-\frac{R^2+a^2-2\,M_{\bullet}\,R}{R^2+a^2}\approx-1+\frac{2\,M_{\bullet}}{R}-\frac{2\,a^2\,M_{\bullet}}{R^3}+...
\end{equation}
Then, we see that the first contribution due to rotation appears at next-to-leading order in the large-distance expansion and may be written as
 \[
 \frac{a^2\,M_{\bullet}}{R^3}=\left(\frac{a}{R}\right)^2\,\left(\frac{M_{\bullet}}{R}\right)\,\,\,;
 \]
since $M_{\bullet}\ll R_g < R$ as discussed before, we may use $\mathsf{a}=\frac{a}{R_g}$ as an expansion parameter. Note that in natural units, this is simply $\mathsf{a}=\frac{v_g}{c}$, where $v_g$ is the average tangential velocity of the galaxy.
 
 \begin{figure}[t]
\centering
\includegraphics[width=0.78\textwidth]{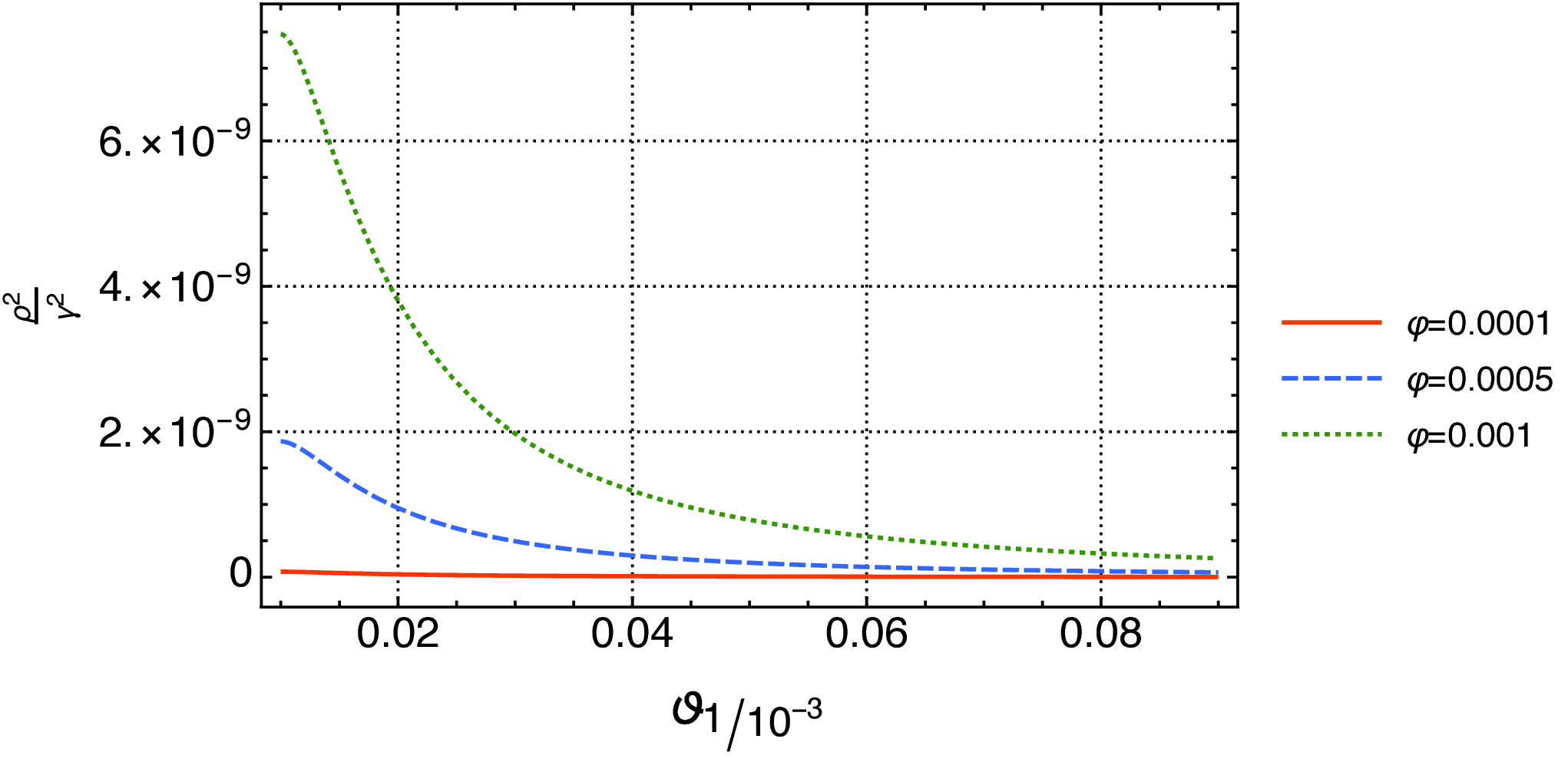}
\caption{A plot of the ratio $\left(\frac{\rho}{\gamma}\right)^2$ as a function of $\vt_1$ for different fixed $\vp$, with the following parameters: $m=10^{12}\,M_{\astrosun}$, $\mathsf{a}=0.0015$, and $R_g=100$ kpc. These are the parameters associated to the Milky Way, previously motivated.}
\label{fig:rhovergamma}
\end{figure}

Thus, the extended nature of a galaxy must be taken into account when describing the distortion it may imprint on a background; however, to consistently use the Kerr metric in this case, one assumes a point-mass approximation, and our map is valid only outside the radius of the galaxy, as stated above. In this sense, the representation in Fig.~\ref{fig:kerrshear} is unphysical: we observe the distorting effect of an object with approximately five times the mass of the Milky Way, which is rotating approximately 20 times faster, at a distance of 100 Mpc from both the observer and lens. Given this setup, the radius of the object spans an angle on the sky of order tens of arc minutes, so the effect represented in Fig.~\ref{fig:kerrshear} lies well within the radius of the object itself.

That being said, it is interesting to note that the computed asymmetrical shearing effect is much larger for an extended object as a rotating galaxy than it is for a black hole. The projected radius of interest beyond which one investigates shearing is typically the \textit{Einstein radius} $\Theta_E$ \cite{Munshi2008} (when the latter is larger than that of the source itself), which for a black hole is approximately the square of the Schwarzschild radius. It is then natural to compare the normalized difference in shear, $\frac{\delta\gamma}{\gamma}$ between the modulus of the shear field on the left- ($\gamma_L$) and on the right-hand ($\gamma_R$) side of the rotation axis at $\Theta_E$ for a black hole and at the galaxy radius $\Theta_R=\frac{R_g}{D_L}$ for a galaxy,
\begin{equation}
\frac{\delta\gamma}{\gamma}=2\,\frac{\gamma_L-\gamma_R}{\gamma_L+\gamma_R}.
\label{deltagamma}
\end{equation}
Considering again a lens at 100 Mpc from both the observer and the background, we find that $\frac{\delta\gamma}{\gamma}$ is of the order $10^{-9}$ for a maximally rotating ($a=1$) $30\,M_{\astrosun}$ black hole and $10^{-6}$ for a super massive $10^9\,M_{\astrosun}$ black hole; yet, $\frac{\delta\gamma}{\gamma}\sim10^{-3}$ for a galaxy with Milky Way parameters at the galaxy radius, which is $\Theta_R\approx10$ arc min in this setup (see Fig.~\ref{fig:deltagamma}). Note that $\Theta_E\sim2\cdot10^{-5}$ arc sec for a $30\,M_{\astrosun}$ black hole and $\Theta_E\sim0.2$ arc sec for a super massive $10^9\,M_{\astrosun}$ black hole.

Let us now briefly discuss whether the $\rho$ field, which is the carrier of the rotational asymmetry and which we identify with a $B$-mode source, might be measurable. Unfortunately, distinguishing the rotation from the shear field may not be experimentally possible, given the resolutions of current and near-future space missions. To gauge our possibilities, we look at simulations of the shear power spectrum in the \textit{Euclid Definition Study Report} \cite{Laureijs2009}. In this case, the region with highest sensitivity is around an amplitude of $10^{-5}\sim10^{-4}$, with a sigma of $0.5\%$. Considering the power spectrum to be roughly constant and averaging over the channels between $l\sim10^3$ and $l\sim10^4$, we estimate that one could whittle down the sigma to $10^{-8}$.
The shear power spectrum will be proportional to the modulus of the shear field squared, $\gamma^2$, and the same goes for the $\rho$ power spectrum, $\rho^2$. For the latter to be distinguishable from the former, $\rho^2$ must be at least comparable to $\sigma_{\gamma^2}$.
To check this, we take the squared ratio of the amplitude of the rotation field $\rho$ and the shear field $\gamma$ obtained through our map, inputting the Milky Way parameters $M_{\rm MW}=10^{12}\,M_{\astrosun}$, $R_{\rm MW}=100$ kpc and $\mathsf{a}_{\rm MW}=1.5\cdot10^{-3}$ relative to our Galaxy -- a plotted example is shown in Fig.~\ref{fig:rhovergamma}. We take $\frac{\rho^2}{\gamma^2}\sim10^{-9}$ as a working value; then, multiplying by the simulated shear power spectrum in Ref.~\cite{Laureijs2009} the rotation power spectrum would be $\rho^2\sim10^{-14}$, which is utterly negligible compared to $\sigma_{\gamma^2}\sim10^{-8}$ obtained above.

Another possible method to extract shear and rotation fields from data is stacking the effect of various spiral galaxies and look for an antisymmetric signal. Taking expected values for the effective number density of galaxies observed by the Large Synoptic Survey Telescope or the \textit{Euclid} survey ($n_{eff} \approx 30$ arc min$^{-2}$ for both \cite{Chang2013, Gressler2009, Laureijs2009}), one can estimate how many foreground galaxies should be stacked in order to measure an asymmetric distortion. Let us consider a foreground galaxy which lenses background galaxies; then, around the lens a circle of radius $\mathsf{r}$ (in arc min) will contain $n_b=n_{eff}\,\pi\mathsf{r}^2$ galaxies. The variance of the shear field $\bs\gamma$, $\sigma_\gamma$, is dominated by the intrinsic ellipticity variance of the lensed sources, which is typically $\sigma_e\approx0.3$ \cite{Munshi2008}. We assume the $\sigma_\rho$ of the rotation field to be roughly the same as $\sigma_\gamma$; then, for a single lens one will be sensitive to perturbations of order $\sigma_\gamma/\sqrt{n_{eff}}$. By stacking the lensing data, one can refine this by a factor $1/\sqrt{n_l}$, where $n_l$ is the number of stacked lenses. Then roughly $\sigma_\gamma\sim\sigma_\rho\sim\frac{0.3}{\mathsf{r}\,\sqrt{n_l\,n_{eff}\,\pi}}\,\,\,$, so in order to get $\sigma_\rho$ of the order of $\rho_{galaxy}$ derivable with our map, one would need to stack $n_l\sim10^{12}$ galaxies for $\mathsf{r}=1$ arc min. Similarly, if one wanted to observe $\delta\gamma = \gamma_L-\gamma_R$, for $\sigma_{\delta\gamma}$ to be of the order of $\delta\gamma_{galaxy}$ one would need to stack $n_l\sim10^8$ lenses. To measure the asymmetric distortion, one would need to know the orientation and rotation velocity of all of these.

\section{Conclusions}
In this paper we have successfully developed a differentiable lensing map in the Kerr space-time and obtained expressions for the shear and rotation fields. This calculation is an instructive result which does not appear in the literature. The action of a Kerr-like object on a background of circular sources is illustrated in Figs.~\ref{fig:kerrshear} and~\ref{fig:kerrvects}, where we have adopted nonphysical parameters to enhance the effect, as it is very small.
    We approximate the metric of rotating galaxies with the Kerr metric (well aware of the caveats of the case) to estimate the possible effect these may have on background radiation. The magnitude of the effect is not expected to be large, but we have quantified it and estimated it for plausible scenarios. Comparing our results with available data, such as Euclid simulations, we find that it is highly unlikely to be able to distinguish the rotation field from the shear field and/or noise. Thus, we conclude that the experimental applications of this map are, as of now, quite limited; we await highly futuristic data to put this work to good use.\\

\paragraph*{Acknowledgements.} We would like to thank Sabino Matarrese for enabling this collaboration. This work was realized during a one year exchange period granted by the Erasmus Mundus program.

\bibliography{KePaBib2}
\end{document}